\documentclass[11pt]{article}
\usepackage{amsmath,amssymb}
\usepackage{framed}

\begin{document}

\setcounter{page}{1}

\pagestyle{plain}

\begin{center}
\Large{\bf Inflation in Energy-Momentum Squared Gravity in Light of Planck2018 }\\
\small \vspace{1cm} {Marzieh
Faraji}$^{}$\footnote{Mari.Faraji@umz.ac.ir},\quad {Narges
Rashidi}$^{}$\footnote{n.rashidi@umz.ac.ir}\quad and
 \quad {Kourosh Nozari}$^{}$\footnote{knozari@umz.ac.ir (Corresponding Author)} \\
\vspace{0.5cm} $^{}$Department of Theoretical Physics, Faculty of
Basic Sciences,
University of Mazandaran,\\
P. O. Box 47416-95447, Babolsar, IRAN\\
\end{center}

\begin{abstract}
We study cosmological dynamics of the energy-momentum squared gravity. By adding the squared of the matter field's energy-momentum tensor ($\zeta\, \textbf{T}^{2}$) to the Einstein Hilbert action, we obtain the Einstein's field equations and study the conservation law. We show that the presence of $\zeta\, \textbf{T}^{2}$ term, breaks the conservation of the energy-momentum tensor of the matter fields. However, an effective energy-momentum tensor in this model is conserved in time. By considering the FRW metric as the background, we find the Friedmann equations and by which we explore the cosmological inflation in $\zeta\,\textbf{T}^{2}$ model. We perform numerical analysis on the perturbation parameters and compare the results with Planck2018 different data sets at $68\%$ and $95\%$ CL, to obtain some constraints on the coupling parameter $\zeta$. We show that \textbf{ for $0< \zeta \leq 2.1\times 10^{-5}$}, the $\zeta\, \textbf{T}^{2}$ gravity is an observationally viable model of inflation.
\end{abstract}

\section{Introduction}

The standard model of cosmology despite all successes in the early
years of development of cosmology, failed to justify some
observations of the universe. Considering a single canonical scalar
field (inflaton) with a flat potential, leading to the slow-roll of
the inflaton and causing the enough exponential expansion of the
early universe, is one simplest way to solve some main problems of
the standard model of cosmology. In the simple single field
inflation model, we get the adiabatic, scale invariant and gaussian
dominant modes of the primordial
perturbations~\cite{Gut81,Lin82,Alb82,Lin90,Lid00a,Lid97,Lyt09,Mal03}.
However, models with non-Gaussian distributed and not exactly scale
invariant perturbations have attracted a lot of
attentions~\cite{Mal03,Bar04,Che10,Fel11a,Fel11b,Noz13a,Noz15,Noz16a,Noz18a,Noz18b,Noz19,Ras20,NoJ11,NoJ18}.

One class of the interesting models in describing the early time
inflation and primordial perturbations is the one related to the
modified gravity. Modified gravity, in its simplest form, is a
function of the Ricci scalar
($f(R)$)~\cite{Sot10,Noj11a,Sta07,Noj16d,Noj16e,Noj16f,Fel10}. A lot
of work on the inflation and perturbations issue, have been done in
the modified gravity and interesting results have been
obtained~\cite{Noj17b,Son07,Noj16c,Noj19a,Odi19,Bud17}. Another
interesting proposal in the modified gravity is to consider an
arbitrary coupling between the Ricci scalar and Lagrangian density
of the matter fields in the
theory~\cite{Ber07,Har08,Tha11,Ber08a,Boe08,Far09,Har10a,Noj04,All05,Ber08b,Far07,Har10b}.
In this regard, some authors have been attracted to the models in
which there is an arbitrary coupling between a function of the Ricci
scalar and the trace of the energy-momentum tensor of the matter
part~\cite{Raj17,Har11,Sah17,Wu18,Paw19}.

In Ref.~\cite{Ros16}, the authors have considered an energy-momentum
squared gravity model in which they have added $\zeta \,
\textbf{T}^{2}$ to the Einstein-Hilbert action which $\textbf{T}^{2}
= T_{\mu \nu}\,T^{\mu \nu} $. In this term, $T_{\mu \nu}$ is the
energy-momentum tensor of the matter part of the theory and $\zeta$
is the coupling constant parameter which its positive values lead to
a  viable cosmological scenario. They have shown that, the presence
of this term leads to a maximum energy density correspond to a
minimum length. In this way there is a bounce in the early universe,
avoiding the early time singularity. Besides, to constraint $\zeta$,
one can find the scalar spectral index, tensor spectral index and
tensor-to-scalar ratio in the model and compare the results with
Planck2018 data set. The constraints on the perturbation parameters
$n_{s}$ and $r$, obtained from Planck2018 TTT, EEE, TTE and EET
data, is as $n_{s}=0.9658\pm0.0038$ and $r<0.072$,
respectively~\cite{Ak18a,Ak18b,Ak18c}. Also, Planck2018 TT, TE, EE
+lowE+lensing+BK14+BAO+LIGO and Virgo2016 data implies the
constraint $-0.62<n_{T}<0.53$ on the tensor spectral
index~\cite{Ak18a,Ak18b,Ak18c}. By using these released data, one
can find some constraints on the model's parameter space.

In this paper, following Ref.~\cite{Ros16}, we consider the
energy-momentum squared gravity model and organize the paper as
follows. In section 2, we obtain the main equations in the $\zeta \,
\textbf{T}^{2}$ gravity model. By considering an additional term as
$\zeta \, \textbf{T}^{2}$ in the Einstein-Hilbert action, we obtain
the Einstein's field equations and study the conservation law for
the effective energy-momentum tensor. We show that the effective
energy-momentum tensor obeys the conservation law. In section 3, we
explore the cosmological dynamics in the $\zeta \, \textbf{T}^{2}$
gravity model. In this regard, we obtain the Friedmann and the
conservation equations of the model. In section 4, the cosmological
inflation in this model is studied. By obtaining the slow-roll
parameters, we find $n_{s}$, $n_{T}$ and $r$ in terms of the model's
parameters. Then, we seek for the observational viability of the
model in confrontation with Planck2018 data. In section 5, we
summarize the model and its results.

\section{The Setup of $\zeta \, \textbf{T}^{2}$ Gravity}
To study the energy-momentum squared gravity model, we start with
the following action
\begin{equation}\label{eq1}
S=\frac{1}{2\kappa}\int\sqrt{-g}\\ \Big[R-2\Lambda-\zeta \,
\textbf{T}^{2}\Big]\,d^4x + S_{M}\,.
\end{equation}
In this action, $g$ is determinant of metric, $R$ is Ricci scalar,
$\kappa =8\pi G$ is the gravitational constant and $\Lambda$ is the
cosmological constant. Note that, as mentioned in~\cite{Ros16}, the
presence of the cosmological constant leads to a positive
acceleration in EMSG model in form of $ \ddot{a} = \frac{2\,
\Lambda}{3} \, a \, , $ without considering any scalar field in the
theory. Also, the matter part of action is defined as follows
\begin{equation}\label{eq2}
S_M=\int\sqrt{-g}\,{\cal{L}}_{M}\,d^4x\,,
\end{equation}
where, ${\cal{L}}_{M}$ is the Lagrangian of the matter fields.
By assuming that the matter Lagrangian density is only a
function of metric and not its derivative, the energy-momentum
tensor of the matter fields is given by
\begin{equation}\label{eq3}
T_{\mu\nu} = \frac {-2}{\sqrt {-g}} \frac {\delta( \sqrt {-g}
{\cal{L}}_{M} )} {\delta g^{\mu\nu}} = {\cal{L}}_{M} g_{\mu\nu} - 2
\frac {\delta {\cal{L}}_{M}} {\delta g^{\mu\nu}}\,.
\end{equation}

By varying action (\ref{eq1}) with respect to the metric
$g^{\mu\nu}$, we obtain the Einstein's field equations in the $\zeta
\, \textbf{T}^{2}$ gravity model as
\begin{equation}\label{eq4}
R_{\mu\nu} - \frac{1}{2} R \,g_{\mu\nu} +\Lambda \,g_{\mu\nu}+ \frac{1}{2} \zeta \, \textbf{T}^{2} \,
g_{\mu\nu} -  \zeta \, \frac{\delta \textbf{T}^{2}}{\delta g^{\mu\nu}} - \kappa\,
T_{\mu\nu} = 0\,.
\end{equation}
While~\cite{Ros16,Nar18}
\begin{equation}\label{eq5}
\frac {\delta \textbf{T}^{2}}{\delta g^{\mu\nu}} = 2 \,\Bigg(T_\mu
^\sigma \, T_{\nu \sigma}+T^{\alpha \beta}\, \frac{\delta  T_{\alpha
\beta}}{\delta g^{\mu \nu}}\Bigg),
\end{equation}
where $ \textbf{T}^2 = T_{\mu \nu}\,T^{\mu \nu}$. Now, we can rewrite equations (\ref{eq4}) as
follows
\begin{equation}\label{eq6}
R_{\mu\nu}-\frac{1}{2} R\,g_{\mu\nu}+ \Lambda \, g_{\mu\nu} =
\kappa\, T_{\mu\nu} - \frac{1}{2} \zeta \, \textbf{T}^2 \,
g_{\mu\nu}+ \zeta \, \Theta _{\mu\nu} + \zeta \, T_{\mu\nu}\,,
\end{equation}
In the above equation, $\Theta_{\mu\nu}$ is defined as
\begin{equation}\label{eq7}
\Theta_{\mu\nu} = -{\cal{L}}_{M}\, g_{\mu\nu} + \frac{1}{2}\,
g_{\mu\nu}\, {\cal{L}}_{M} - \frac {1} {2} T_{\mu\nu} - 2\,
g^{\mu\nu}\, \frac {\partial^2 {\cal{L}}_{M}}{\partial g^{\mu\nu}
\partial g^{\alpha\beta}}.
\end{equation}
Considering that all the changes appear in the right hand side of
field equations, we can introduce the following effective
energy-momentum tensor
\begin{equation}\label{eq8}
T^{eff}_{\mu\nu} = T_{\mu\nu} + \frac {1}{\kappa}
\, \zeta \, \Theta_{\mu\nu} +\zeta \, T_{\mu\nu} - \frac {1}{2}
g_{\mu\nu} \, \zeta \, \textbf{T}^{2}.
\end{equation}
and rewrite the Einstein field equations as follows
\begin{equation}\label{eq9}
G_{\mu\nu} + \Lambda \, g_{\mu\nu}= \kappa\, T_{\mu\nu}^{eff}.
\end{equation}
Note that, these equations are also obtained in Ref.~\cite{Boa17},
where the authors have considered $\zeta \,(T^{\mu\nu}\,T_{\mu\nu})^n $.
In this work, we study the case with $n=1$.

Now, we seek for the conservation law in this model. In this regard,
we obtain the covariant derivative of the energy-momentum tensor as
\begin{eqnarray}\label{eq11}
\nabla^{\mu}\, T_{\mu\nu} =  \frac {1} {(\kappa +
\frac{1}{2}\,\zeta)} \Bigg[ \frac {1}{2} \,\zeta \, \textbf{T}^{2}\,
g_{\mu\nu} + {\cal{L}}_{M}\, \zeta \, g_{\mu\nu}\nonumber\\ - \frac
{1}{2}\, \zeta \, g_{\mu\nu}\, {\cal{L}}_{M} + 2 \, \zeta \,
g^{\alpha\beta} \frac {\partial^2 {\cal{L}}_{M}} {\partial
g^{\mu\nu}\partial g^{\alpha\beta}} \Bigg].
\end{eqnarray}
This equation shows that by adding $\zeta \, \textbf{T}^{2}$ to the
action, the conservation law breaks. This means that, due to the
interaction between the matter and curvature sectors, the
energy-momentum tensor is not conserved in time. However, we have
introduced an effective energy-momentum tensor in equation
(\ref{eq9}) which satisfies the conservation equation. In fact, we
have
\begin{equation}\label{eq12}
\nabla^{\mu}\, T^{eff}_{\mu\nu} = 0\,,
\end{equation}
which shows that the effective energy-momentum tensor is conserved
in time.

Up to here, we have obtained the main equations of the
energy-momentum squared gravity model and studied the conservation
law in this model. In the next section, we explore the cosmological
implications of this interesting model.

\section{Cosmology in $\zeta \, \textbf{T}^{2}$ Gravity}
Since our universe is homogenous and isotropic in large scales, it
is convenient to adopt the following Friedmann-Robertson-Walker
metric as the background
\begin{equation}\label{eq13}
ds^2 = dx^{\mu} dx^{\nu} g_{\mu \nu} = - dt^2 + a(t)^2 \bigg[ \frac
{dr^2}{1-k r^2} + r^2 (d\theta^2 + sin^2 \theta d\phi^2)\bigg].
\end{equation}
Considering that the observational data confirms the flat universe,
we consider the FRW metric with $k=0$. To obtain the Friedmann
equations, it is necessary to adopt a suitable choice for the
Lagrangian since the tensor $\Theta _{tt}$ is related to the matter
field's Lagrangian via equation (\ref{eq7}). It has been shown that
one can choose either ${\cal{L}}_M=p$ (where $p$ is the pressure) or
${\cal{L}}_M=-\rho$ (where $\rho$ is the energy density) in
Refs.~\cite{Sot02,Ber08}. Here, we adopt the case ${\cal{L}}_M=p$.
In this regard, we find
\begin{equation}\label{eq14}
H^2= \frac{\kappa}{3} \, \rho + \frac{\Lambda}{3}\,- \zeta
\,(\frac{1}{2} \, p^2 \, + \frac{4}{3} \, \rho \, p + \frac{1}{6} \,
\rho^2)
\end{equation}

\begin{equation}\label{eq15}
\frac {\ddot a}{a} = -\frac{\kappa}{6} \, (\rho+3\,p)+
\frac{\Lambda}{3} + \zeta \, (p^2 + \frac{2}{3} \, \rho \, p +
\frac{1}{3} \rho^2)
\end{equation}
As we mentioned $\zeta$ is the coupling constant and the important point about it is that,
against other coupling constants in higher order theories of gravity
which are dimensionless, dimension of $\zeta$ is
$\frac{s^{4}}{kg^{2}}$ (Note that, in this paper we adopt the light
speed as $c=1$). Further more as mentioned in Ref.~\cite{Ros16} this
coupling constant should be both positive and small enough in order
to give interesting cosmological results and pass classical
gravitational tests in low energy regimes respectively. We see in
the next section that both of these conditions are satisfied in our
observational analysis.

From equations $(13)$ and $(14)$, we can define the
following effective energy density and pressure
\begin{equation}\label{eq18}
\rho_ {eff} = \rho - \frac {1}{2} \zeta\, \Big(\rho ^2 + 3p^2
+ 8 \rho\, p \Big)\,,
\end{equation}

\begin{equation}\label{eq19}
p_ {eff}= p-\frac {1} {2} \zeta\, \Big(\rho^2 + 3p^2\Big)\,,
\end{equation}
leading to the following continuity equation
\begin{equation}\label{eq20}
{\dot{\rho}}_{eff} + 3 \, H (\rho_{eff} + p_{eff}) \, =0
\end{equation}
Which dot means derivatives with respect to cosmological time. This
equation means that the effective energy density is conserved in
time.

After obtaining the main cosmological equation in the $\zeta\,
\textbf{T}^2 $ gravity model, we study the inflation and
observational viability of this model in the next section.

\section{Inflation and Observational Viability of $\zeta\, \textbf{T}^2$ Gravity}

By using the definition of a constant and positive equation of state
as $\omega = \frac {p} {\rho}$ \,, we can rewrite the Friedmann
equations (\ref{eq14}) and (\ref{eq15}) as follows respectively
\begin{equation}\label{eq21}
H^2= \frac {\kappa} {3}\, \rho  + \frac{\Lambda}{3} - \frac {1} {6}\, \lambda \,\rho^2\,,
\end{equation}
\begin{equation}\label{eq22}
\dot H + H^2 = - \frac{\kappa}{6} \, \xi \, \rho  + \frac{\Lambda}{3} + \beta \, \rho^2
\end{equation}
where constants are defined as
\begin{equation}\label{eq23}
\lambda = \zeta (3 \, \omega^2 + 8 \, \omega + 1) \,.
\end{equation}
\begin{equation}\label{eq24}
\xi = 1 + 3 \, \omega
\end{equation}
\begin{equation}\label{eq25}
\beta = \zeta \, (\omega^2 \, + \frac{2}{3} \, \omega + \frac{1}{3} )
\end{equation}
By differentiating the equation $(19)$ with respect to cosmological
time, we obtain
\begin{equation}\label{eq26}
\ddot H + 2 H \dot H = - \frac{\kappa}{6} \xi\, \dot{\rho} + 2 \beta \,\rho
\,\dot{\rho}\,.
\end{equation}
Which $\rho$ and $\dot{\rho}$ must be substitute.
By rearranging the equation $(18)$ in order to express the energy density, we have
\begin{equation}\label{eq27}
\rho = \frac{\kappa + \sqrt{-6 \, H^2 \, \lambda+ 2 \, \Lambda \,
\lambda + \kappa ^2}}{\lambda}
\end{equation}
Also, the time derivative of the energy density by using equations $(15)$, $(16)$ and $(17)$ is given by
\begin{eqnarray}\label{eq28}
\dot \rho = \frac{H}{\lambda ^2 \, (-1 + \kappa + \sqrt{-6 \, H^2 \,
\lambda + 2 \, \Lambda \, \lambda + \kappa ^2} )} \bigg[9H^2 \,
\zeta \, \xi \, \lambda + 9H^2 \, \lambda^2 \nonumber\\- 3 \,\zeta
\,\xi \, \kappa \sqrt{-6 \, H^2 \, \lambda + 2 \, \Lambda \, \lambda
+ \kappa ^2}\, - 3 \, \xi \, \zeta \, \kappa^2 -3 \, \zeta \, \xi \,
\lambda \, \Lambda - 3 \,\kappa \, \lambda \sqrt{-6 \, H^2 \,
\lambda + 2 \, \Lambda \, \lambda + \kappa ^2} \nonumber\\+ 3 \,
\chi \, \lambda \, \sqrt{-6 \, H^2 \, \lambda + 2 \, \Lambda \,
\lambda + \kappa ^2} - 3 \, \kappa^2 \, \lambda + 3 \, \chi \,
\lambda
  \, \kappa - 2 \, \Lambda \, \lambda^2\bigg]
\end{eqnarray}
Where we have defined
\begin{equation}\label{eq30}
\chi = \zeta \, \xi\,,
\end{equation}
By substituting equations $(24)$ and $(25)$ into equation
$(23)$, we obtain a Li$\acute{e}$nard equation as
\begin{equation}\label{eq33}
\ddot H + A H \dot H + B H^3 = 0\,.
\end{equation}
Where the parameters $A$ and $B$ are
\begin{equation}\label{eq34}
A = \frac{4 \, \zeta}{\kappa} \,\,,\quad\quad  B= \frac {- 27 \,
\zeta} {\kappa} + \frac{2 \, \Lambda}{\kappa^2}\,.
\end{equation}
In Ref.~\cite{Ros16} it has been shown that, to resolve the
singularity of the early universe and achieve the positive
acceleration expansion in the energy-momentum squared gravity model the universe should be
dominated with the radiation component. In this regard, to obtain
the coefficients (\ref{eq34}), we have adopted $\omega=\frac{1}{3}$.
The solution of the differential equation (\ref{eq33}) in the
general form is
\begin{equation}\label{eq35}
H = \bigg( \frac {-2z}{A}\bigg)^ {\frac {1}{2}}\,,
\end{equation}
which is given in the parametric form. Also, the parameter $z$ is
given by
\begin{equation}\label{eq36}
z={\cal{C}} \exp\bigg( - \int {\frac {\sigma d\sigma}{ \sigma^2 -
\sigma - D } } \bigg)\,,
\end{equation}
where ${\cal{C}}$ is a constant and we have
\begin{equation}\label{eq37}
D= \frac {-2B}{A^2}\,.
\end{equation}
To see more details about obtaining the solution of the differential
equation (\ref{eq33}), see the Appendix and also Ref.~\cite{Pol03}.
Now, we can obtain the derivatives of $H$ as follows
\begin{equation}\label{eq38}
\dot H = \sigma z
\end{equation}
\begin{equation}\label{eq39}
\ddot H = - A \,(\frac{\sigma + D}{\sigma})\, H \dot H
\end{equation}
and
\begin{equation}\label{eq40}
\dddot H = \dot H \Bigg[A^2 H^2  \bigg(\frac {\sigma + D}
{\sigma}\bigg)^2 - A \dot H \bigg(\frac {\sigma +D
}{\sigma}\bigg)\Bigg]\,.
\end{equation}
We can use the equations (\ref{eq35}) and (\ref{eq38})-(\ref{eq40})
to study the inflation in the $\zeta\, \textbf{T}^2 $ gravity model. To this
end, we define the following slow-roll parameters~\cite{Mar14}
\begin{equation}\label{eq41}
\epsilon_{1} \equiv \frac {- \dot H} {H^2}\,,
\end{equation}
\begin{equation}\label{eq42}
\epsilon_{2} \equiv \frac {\ddot H}{H \dot H} - \frac {2 \dot
H}{H^2}\,,
\end{equation}
and
\begin{equation}\label{eq43}
\epsilon_{3} \equiv \bigg(\ddot H H - 2 \dot H^2\bigg)^{-1}\, \Bigg[
\frac { H \dot H \dddot H - \ddot H (\dot H^2 + H \ddot H )} {H \dot
H } - \frac {2 \dot H}{H^2} \Big(H \ddot H - 2 \dot H^2 \Big)
\Bigg]\,.
\end{equation}
It is possible to write the slow-roll parameters in terms of the
e-folds number. The e-folds number is defined as
\begin{equation}\label{eq44}
N\equiv \ln \bigg(\frac {a_{f}}{a_{i}}\bigg) = - \int_{a_{i}}
^{a_{f}} H(t)\, dt\,,
\end{equation}
where $a_{i}$ and $a_{f}$ show the values of the scale factor at the
beginning and end of the inflation era, respectively. The e-folds
number in $\zeta\, \textbf{T}^2 $ gravity model takes the following
form
\begin{equation}\label{eq45}
N = \frac {2} {A \sqrt {4D+1}} \tanh^{-1} \bigg(\frac {2\sigma
+1}{\sqrt {4D+1}}\bigg).
\end{equation}
By using the definition of the e-folds number given in equation
(\ref{eq44}), we obtain the following expressions for the slow-roll
parameters
\begin{equation}\label{eq46}
\epsilon_{1}(N)\equiv \frac {-H^\prime (N)} {H(N)}\,,
\end{equation}
\begin{equation}\label{eq47}
\epsilon_{2} (N) \equiv \frac {H^{\prime \prime} (N_)}{H ^\prime (N)
} - \frac {H^\prime (N)} {H(N)}\,,
\end{equation}
and
\begin{equation}\label{eq48}
\epsilon_{3}(N) \equiv \Bigg[\frac {H(N) H^{\prime}(N) }{H^ {\prime
\prime} (N) H(N) - H ^{\prime 2}(N) }\Bigg] \Bigg[ \frac { H^
{\prime \prime \prime }(N) } { H ^{\prime} (N) } - \frac { H
^{\prime \prime 2 } (N)} { H^{\prime 2} (N) }- \frac { H^{\prime
\prime}(N) } {H(N)} + \frac { H ^{\prime 2} (N)} {H^2(N)} \Bigg]\,.
\end{equation}
These parameters in our $\zeta\, \textbf{T}^2 $ model, take the following
form
\begin{equation}\label{eq49}
\epsilon_{1}= \frac {\sigma A}{2}\,,
\end{equation}
\begin{equation}\label{eq50}
\epsilon_{2} = A \bigg(\frac {\sigma ^2 -\sigma -
D}{\sigma}\bigg)\,,
\end{equation}
\begin{equation}\label{eq51}
\epsilon _{3} = A \bigg( \frac {- \sigma ^3 + \sigma + D} {-\sigma
^2 + \sigma +D }\bigg)\,.
\end{equation}

Now, by considering the equations (\ref{eq45}) and
(\ref{eq49})-(\ref{eq51}), we can seek for graceful exit of the
model from inflation era. In the inflation era we have
$\epsilon_{1},\epsilon_{2},\epsilon_{3}\ll 1$. To have graceful exit
from inflation, one of the slow-roll parameters should reach unity.
In this regard, we plot the parameters $\epsilon_{1}$ and
$\epsilon_{2}$ versus the e-folds number for two sample values of
$\zeta$. The results are shown in figure 1. As figure shows, the
slow-roll parameter $\epsilon_{2}$ meet unity at $N=60$. This means
that in our model inflation ends after 60 e-folding.

\begin{figure}
\begin{center}\includegraphics{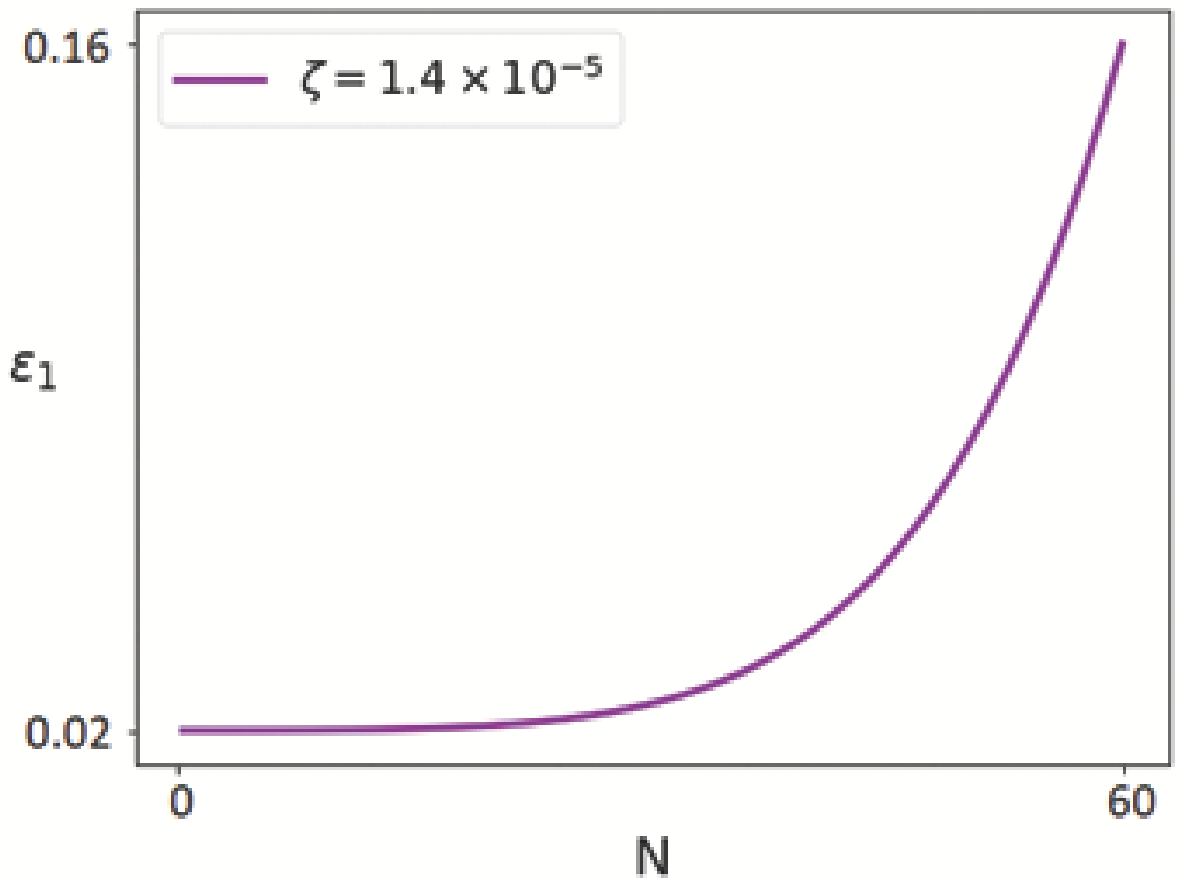}\includegraphics{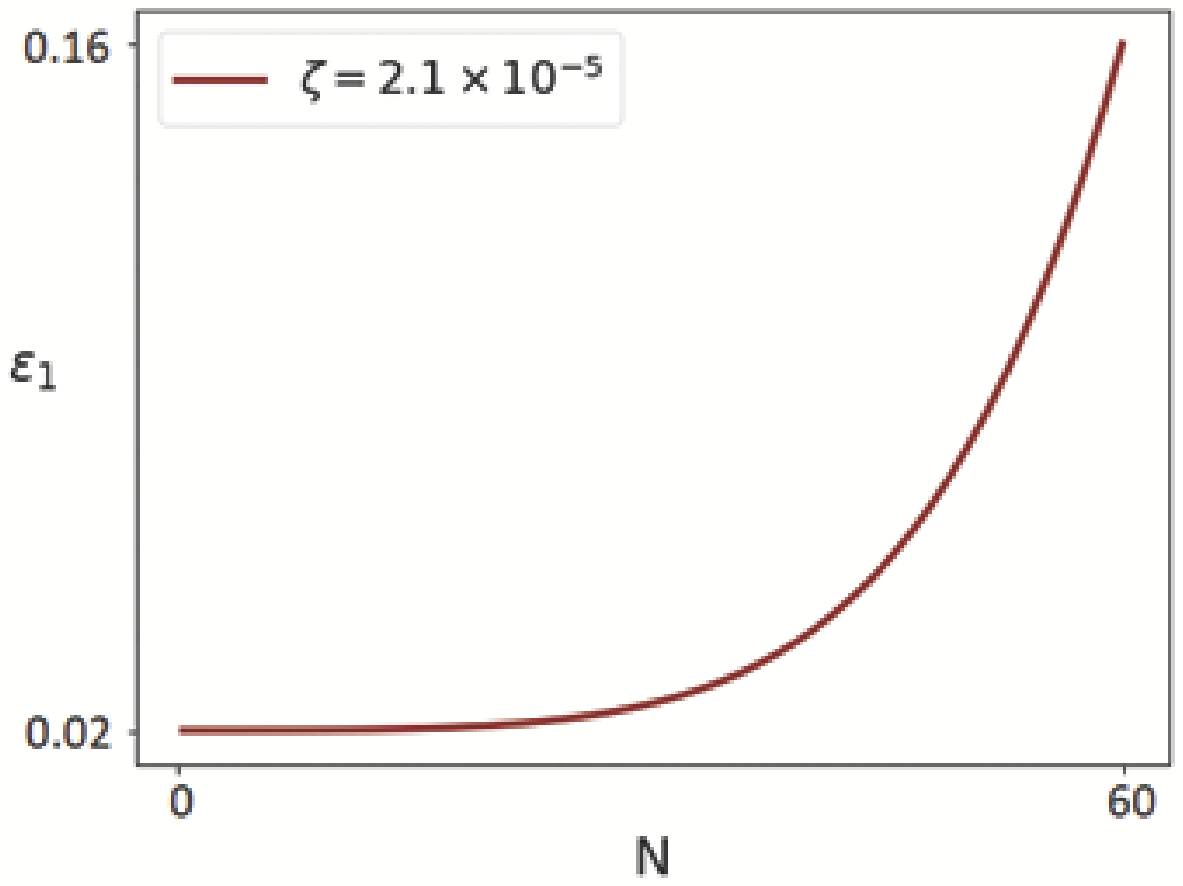}\vspace{6cm}
\includegraphics{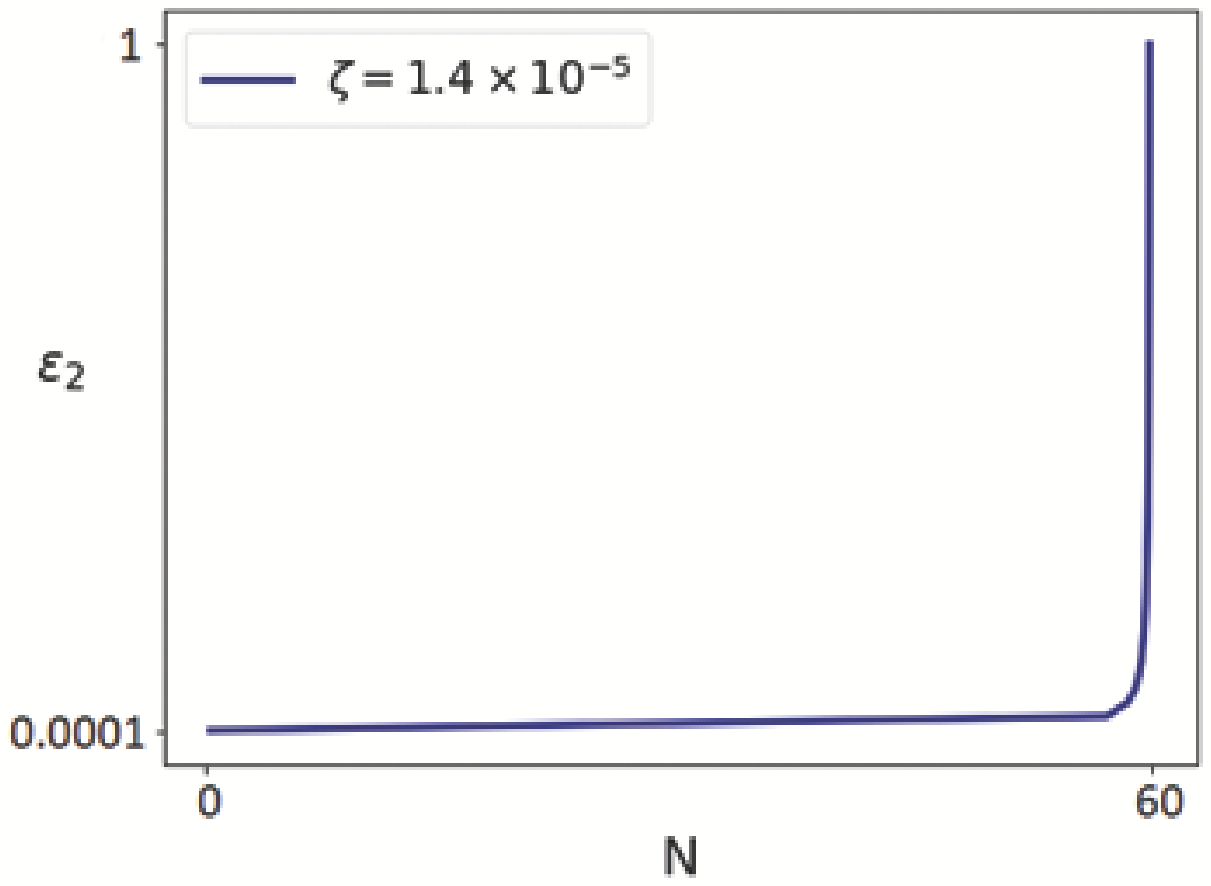}\includegraphics{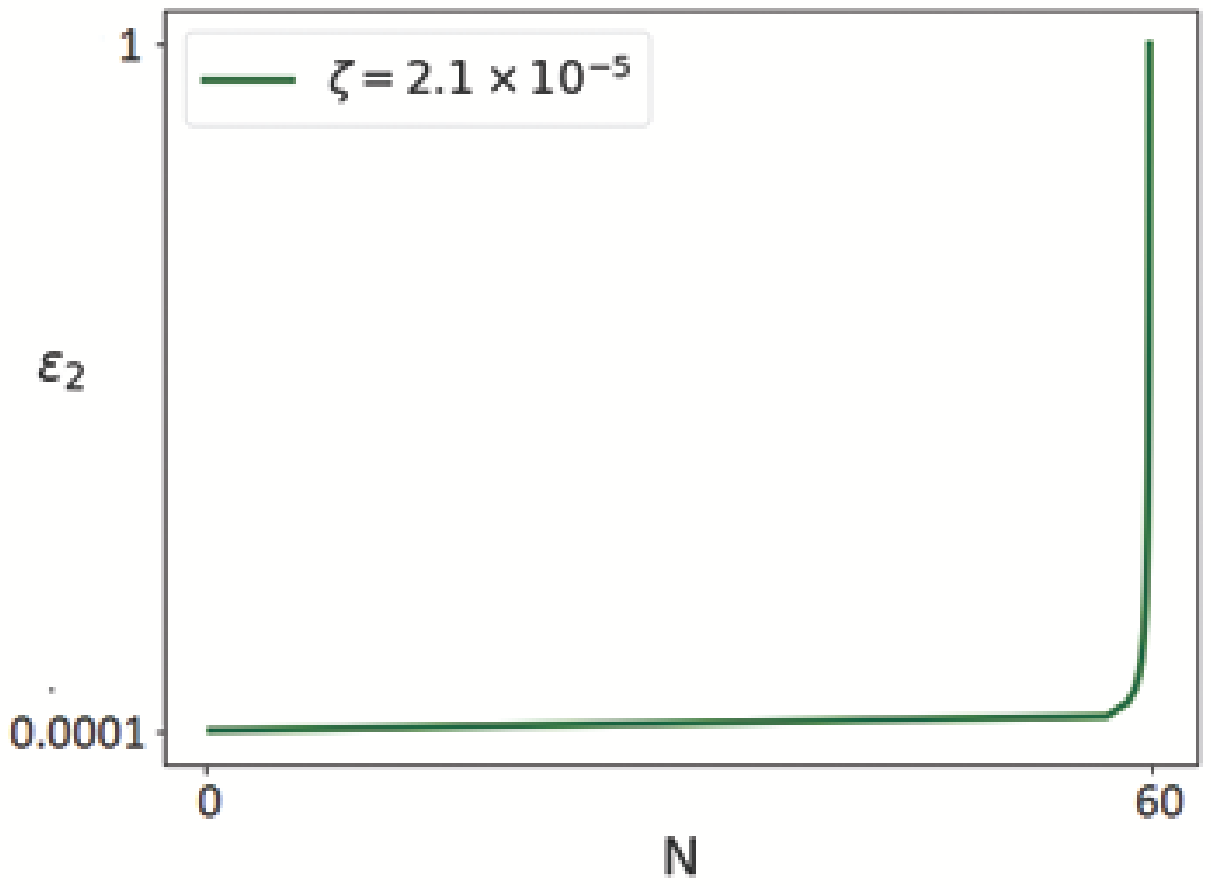}\vspace{6cm}
\end{center}
\caption{\label{fig1}\small {The evolution of the slow-roll
parameters $\epsilon_{1}$ and $\epsilon_{2}$ versus the e-folds
number during the inflation for the two sample values of $\zeta$.}}
\end{figure}

Another way to seek for inflation and its a graceful exit is the
study of the evolution of the Hubble parameter versus the e-folds
number. The result is shown in figure 2, which has been plotted for
$\zeta=2.1\times 10^{-5}$. As this figure shows, the Hubble
parameter during the inflation changes very slowly until inflation
ends.

\begin{figure}
\begin{center}\includegraphics{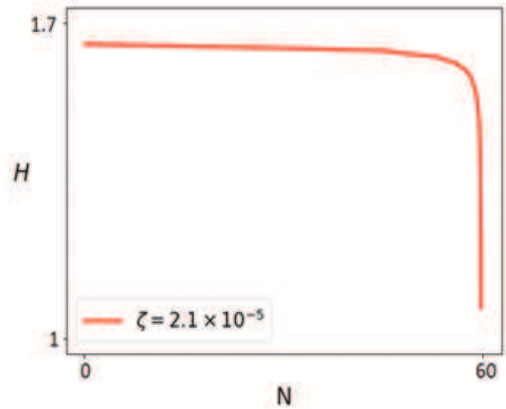}\vspace{6cm}
\end{center}
\caption{\label{fig3}\small {The evolution of the Hubble parameter
versus the e-folds number during inflation.}}
\end{figure}

Also, the perturbation parameters are defined in terms of the
slow-roll parameters. In this regard, the scalar spectral index and
its running are given by ~\cite{Noj16f,Raj17,Mar14,Bam14,Bha20}
\begin{equation}\label{eq52}
n_{s} \approx 1-2\epsilon_{1} - 2\epsilon_{2}\,,
\end{equation}
and
\begin{equation}\label{eq53}
\alpha_{s} \approx - 2\epsilon_{1} \epsilon_{2} - \epsilon_{2}
\epsilon_{3}\,.
\end{equation}
respectively. The tensor spectral index, in terms of the slow-roll
parameters, is defined as
\begin{equation}\label{eq54}
n_{T} \approx -2\epsilon_{1}.
\end{equation}
Finally, the tenor-to-scalar ratio is given by
\begin{equation}\label{eq55}
r\approx 16 \epsilon _{1}\,.
\end{equation}

By substituting equations (\ref{eq49})-(\ref{eq51}) in equations
(\ref{eq52})-(\ref{eq55}) we obtain the following expressions for
the perturbation parameters
\begin{equation}\label{eq56}
n_{s} = 1 - A \sigma - 2\, A\, \bigg(\frac {\sigma^2 - \sigma -
D}{\sigma}\bigg)\,,
\end{equation}
\begin{equation}\label{eq57}
\alpha_{s} = \Bigg[- A^2\, \bigg(\frac {\sigma ^2 - \sigma -
D}{\sigma}\bigg)\Bigg]\Bigg[\sigma +  \frac {- \sigma ^3 + \sigma +
D}{-\sigma ^2 +\sigma +D}\Bigg]\,,
\end{equation}
\begin{equation}\label{eq58}
n_{T} = - A\,\sigma\,,
\end{equation}
and
\begin{equation}\label{eq59}
r =  8 \,A \,\sigma\,.
\end{equation}
Finally, by using equation (\ref{eq59}) to eliminate the parameter
$\sigma$, we get
\begin{equation}\label{eq60}
n_{s} = \frac {3}{4} - \frac {r} {8} + 2A - \frac {32 B}{r}.
\end{equation}
\begin{equation}\label{eq61}
\alpha _{s} = \left(\frac {r} {8A} -1 + \frac {16B}{AR}\right)
\left[\frac {-rA}{8} - A^2 (\frac {-r^3 + 64 r A^2 - 1024 AB}{-r^2 +
8 Ar - 128 B})\right].
\end{equation}
and
\begin{equation}\label{eq62}
n_{T} = \frac {-r}{8}.
\end{equation}

After obtaining the main perturbations parameters, now we explore
the model numerically and compare the results with the observational
data. In this regard, we can examine the observational viability of
our setup and obtain some constraints on the coupling parameter
$\zeta$. Note that, in Ref.~\cite{Ros16} it has been shown that only
positive values of $\zeta$ lead to the viable cosmology. Therefore,
in our analysis, we consider only the positive values of this
parameter. Figure 3 shows the behavior of the running of the scalar
spectral index versus the scalar spectral index, in the background
of the Planck2018 TT, TE, EE+lowE+lensing data~\cite{Ak18a}.
 To plot this figure, we have used equations (\ref{eq60}) and
(\ref{eq61}), where the parameters $A$ and $B$ are given by equation
(\ref{eq34}). This figure and forthcoming figures have been plotted
for $0< \zeta \leq 10^{-5}$. We can also study the behavior of the
tensor-to-scalar ratio versus the scalar spectral index by using
equation (\ref{eq60}). The result is shown in figure 4, in the
background of the Planck2018 TT, TE, EE+lowE+lensing+BK14+BAO data
set~\cite{Ak18b}. Also, figure 5 shows the tensor-to-scalar ratio
versus the tensor spectral index (see equation (\ref{eq62})) in the
background of the Planck2018 TT, TE, EE+lowE+lensing+BK14+BAO+LIGO
and Virgo2016 data~\cite{Ak18b}. To plot figures 3-5, we have
borrowed the contour plots released by Planck 208
team~\cite{Ak18a,Ak18b,Ak18c}. This is because, in this paper, we
compare the results of the numerical analysis in our model with the
Planck observational data. However, the blue regions are the
numerical results of our setup which have been obtained from
equations (\ref{eq60})-(\ref{eq62}). As these figures show, the
energy-momentum squared gravity model in some ranges of the model's
parameter space is consistent with observational data. By performing
the numerical analysis, we have obtained some ranges of the
parameter $\zeta$ which cause the viability of the model in
confrontation with different data sets. The constraints are
summarized in table 1.

\begin{figure}
\begin{center}\includegraphics{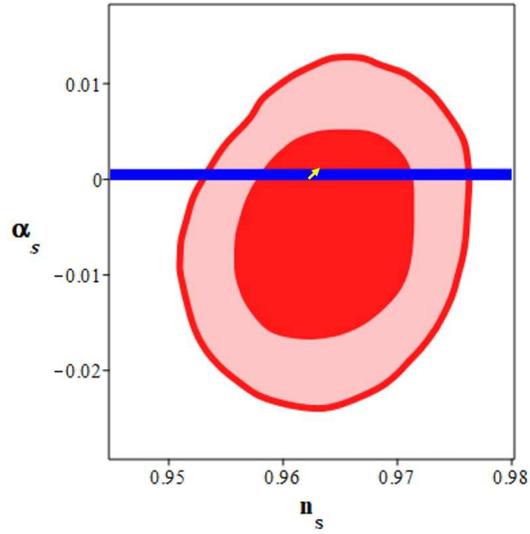}\vspace{6cm}
\end{center}
\caption{\label{fig3}\small {Running of the scalar spectral index
versus the scalar spectral index for $0< \zeta \leq 10^{-5}$ (blue
region), in the background of Planck2018 TT, TE, EE+lowE+lensing
data (red regions). The yellow arrow shows the direction in which
the parameter $\zeta$ increases.}}
\end{figure}

\begin{figure}
\begin{center}\includegraphics{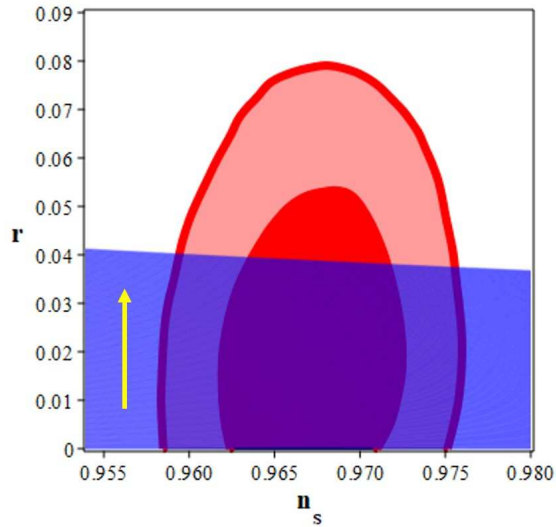}\vspace{6.5cm}
\end{center}
\caption{\label{fig4}\small {Tensor-to-scalar ratio versus the
scalar spectral index for $0< \zeta \leq 10^{-5}$ (blue region), in
the background of the Planck2018 TT, TE, EE+lowE+lensing+BK14+BAO
data set (red regions). The yellow arrow shows the direction in
which the parameter $\zeta$ increases.}}
\end{figure}

\begin{figure}
\begin{center}\includegraphics{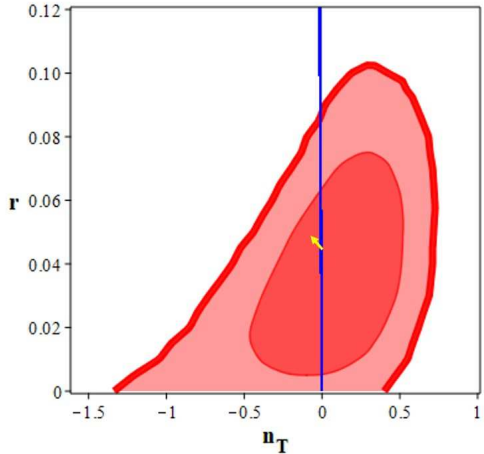}\vspace{6cm}
\end{center}
\caption{\label{fig5}\small {Tensor spectral index versus the tensor
spectral index for $0< \zeta \leq 10^{-5}$ (blue region), in the
background of the Planck2018 TT, TE, EE+lowE+lensing+BK14+BAO+LIGO
and Virgo2016 data (red regions). The yellow arrow shows the
direction in which the parameter $\zeta$ increases.}}
\end{figure}

\begin{table*}
\tiny\caption{\small{\label{tab:1} The ranges of the parameter
$\zeta$ $\left(\frac{s^{4}}{kg^{2}}\right)$ in which the
tensor-to-scalar ratio, the scalar spectral index, its running and
the tensor spectral index of the $\zeta\, \textbf{T}^2 $ gravity model are
consistent with different data sets.}}
\begin{center}
\begin{tabular}{ccccccc}
\\ \hline \hline \\ && Planck2018 TT, TE, EE+lowE& Planck2018 TT, TE, EE+lowE&Planck2018 TT, TE, EE+lowE&
\\
&$$& +lensing & +lensing+BK14+BAO&lensing+BK14+BAO&
\\
&$$&  & &+LIGO$\&$Virgo2016 &
\\
\hline \\ &$68\%$ CL&$  \zeta \leq
1.6\times 10^{-3}$ &$ 0<\zeta \leq 1.4\times 10^{-5}$&$0<\zeta$&\\
\\ \hline\\ \\
&$95\%$ CL&$ 0<\zeta \leq 1.6\times 10^{-3}$&$0< \zeta \leq
2.1\times
10^{-5}$&$0<\zeta$&\\ \\
\hline
\end{tabular}
\end{center}
\end{table*}

After studying the perturbation's parameters numerically and
obtaining some constraints on the model from the observational data,
it seems interesting to seek the abundance of the the fluid $\rho$
with $\omega=\frac{1}{3}$ (corresponding to radiation component).
For this purpose, we rewrite equation (\ref{eq21}) in terms of the
density parameters $\Omega$ as
\begin{equation}\label{eq63}
1=\Omega_{{rad}}+\Omega_{{\Lambda}}-{\Omega^{2}_{{rad}}}\Omega_{{\zeta}}\,,
\end{equation}
Where
\begin{equation}\label{eq64}
\Omega_{{rad}}={\frac {\kappa\,\rho}{3{H}^{2}}}\,,\quad
\Omega_{{\Lambda}}={\frac {\Lambda}{3{H}^{2}}}\,,\quad
\Omega_{{\zeta}}={\frac {3\lambda\,{H}^{2}}{2{\kappa}^{2}}}\,,
\end{equation}
With $rad$ presenting the radiation component. As our
numerical analysis has shown, the strength of the energy-momentum
squared gravity in our model is small ($0< \zeta \leq 2.1\times
10^{-5}$). This means that even for small strength of the
energy-momentum squared gravity, it is possible to get the
observationally viable inflationary model. In this sense, to study
the abundance of the the fluids in our model, we adopt small value
of $\Omega_{\zeta}$ as $\Omega_{\zeta}=0.001$. Then, we find the
abundance of $\Omega_{r}$ and $\Omega_{\Lambda}$ at $68\%$ CL and
$98\%$ CL, for this adopted value of $\Omega_{\zeta}$. The result is
shown in figure 6. According to our analysis at $68\%$ CL, we have
$\Omega_{r}=0.908\pm 0.003$ and $\Omega_{\Lambda}=0.091\pm 0.003$.

\begin{figure}
\begin{center}\includegraphics{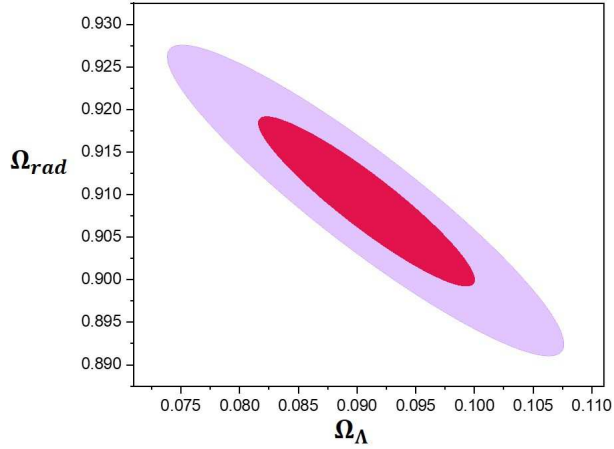}\vspace{6cm}
\end{center}
\caption{\label{fig6}\small {The abundance of $\Omega_{r}$ and
$\Omega_{\Lambda}$ at $68\%$ CL (red region) and $98\%$ CL (plum
region), for $\Omega_{\zeta}=0.001$.}}
\end{figure}

\section{Conclusion}
In this paper, we have studied the cosmological dynamics of the
energy-momentum squared gravity. In this regard, we have considered
an additional term in the Einstein-Hilbert action as $ \zeta\,
\textbf{T}^2 $, where $\zeta$ is a positive constant coupling and
$\textbf{T}^2 = T_{\mu\nu}\, T^{\mu\nu} $. We have presented the
Einstein's field equations in $\zeta\, \textbf{T}^2$ gravity model
and also studied the conservation law via the energy-momentum
tensor. We have shown that, by adding a $\zeta\, \textbf{T}^2$ term
to the action, the energy momentum of the matter fields breaks.
However, if we consider an effective energy-momentum tensor, the
conservation law would be satisfied. After that, by assuming the FRW
metric as the background, we have obtained the Friedmann equations
in this setup. In this regard, we have introduced the effective
energy density and the effective pressure, by which we have shown
the conservation of the effective energy density in the energy-momentum squared gravity model.

Then, we have studied the inflation phase in this model. By
obtaining the slow-roll parameters in this model, we have expressed
the perturbation parameters in terms of the model's parameters. By
performing a numerical analysis on the scalar spectral index, its
running, tensor spectral index and tensor-to-scalar ratio, we have
studied the viability of the $\zeta\, \textbf{T}^2 $ model in the context of
the inflation. according to our analysis, Planck2018 TT, TE,
EE+lowE+lensing+BK14+BAO+LIGO and Virgo2016 data doesn't give any
constraint on the coupling parameter $\zeta$. However, it is
possible to set constraints on $\zeta$ by considering the Planck2018
TT, TE, EE+lowE+lensing and also, Planck2018 TT, TE,
EE+lowE+lensing+BK14+BAO data sets.

In summary, the $\zeta\, \textbf{T}^2 $ gravity model, which lead to
the bounce at early universe, is an observationally viable inflation
model with $0< \zeta \leq 2.1\times 10^{-5}$.\\

\textbf{Appendix}\\ \\We have the following Lienard differential
equation
$$
\ddot H + A H \dot H + B H^3 = 0\,,
$$
where
$$
A = \frac{4 \, \zeta}{\kappa}
$$
and
$$
B = \frac {- 27 \, \zeta} {\kappa}+ \frac{2 \, \Lambda}{\kappa^2}\,
$$
By assuming $ H \equiv y $, we can rewrite the above Lienard
equation as follows
$$
\ddot y + f(y) \dot y + g(y) =0\,,
$$
where
$$
f(y) = A \, y \,,
$$
and
$$
g(y) = B \, y^3 \,.
$$
Now, we define $w(y) = \dot y$ and by which we convert our Lienard
equation to the Abel differential equation of the second kind as
$$
w w' + f(y) w + g(y) = 0\,,
$$
with $w w' = \ddot y$ and $ w' \equiv \frac {dw}{dy}$. By
introducing $z=\int F(y) dy$, with $ F(y) = -A y$ and $ G(y) = - B
y^3$, the Abel equation takes the following canonical form
$$
w w_{,z} = w + \phi (z)\,,
$$
where $``,z $" demonstrates derivative with respect to $z$, and
$$
\phi(z) = \frac {G(y)} {F(y)}\,.
$$
Considering that
$$
z = - \frac {A y^2}{2}\,,
$$
we find the following expression for $\phi(z)$
$$
\phi (z) = \frac {B}{A}\, y^2 \,,
$$
leading to
$$
w w_{,z}=w + \frac {B}{A} \, y^2 =w+D\,z\,.
$$
By defining $k(z)=Dz$, we get
$$
w w_{,z}=w+k(z)\,,
$$
which has the following solution
$$
z={\cal{C}} \exp\bigg( - \int {\frac {\sigma d\sigma}{ \sigma^2 -
\sigma - D } } \bigg)\,,\quad w={\cal{C}}\sigma \exp\bigg( - \int
{\frac {\sigma d\sigma}{ \sigma^2 - \sigma - D } } \bigg)\,.
$$
Now, we can obtain the Hubble parameter and its derivatives. From
$y\equiv H$ and $z=-\frac{Ay^2}{2}$, we find
$$
H = \left(\frac {-2z}{A}\right)^{\frac{1}{2}}\,.
$$
Using $\dot{H}=\dot{y}=w$, we obtain
$$
\dot H = \sigma\,z\,.
$$
Also, from $\ddot H = w w'$ and considering that
$\frac{dw}{dy}\equiv\frac{dw}{dz}\frac{dz}{dy}$, we get
$$
\ddot H =  -A \left(\frac {\sigma + D}{\sigma}\right) H \dot H\,.
$$


\begin{thebibliography}{dummy}

\bibitem{Gut81} Guth, A., Phys. Rev. D, \textbf{23}, 347 (1981).

\bibitem{Lin82} Linde, A. D, Phys. Lett. B, \textbf{108}, 389 (1982).

\bibitem{Alb82} Albrecht A., \& Steinhard, P., Phys. Rev. D, \textbf{48}, 1220 (1982).

\bibitem{Lin90} Linde, A. D. 1990, \emph{Particle Physics and Inflationary Cosmology}
(Harwood Academic Publishers, Chur, Switzerland).

\bibitem{Lid00a}Liddle, A. \& Lyth, D. 2000, \emph{Cosmological Inflation and Large-Scale Structure},
(Cambridge University Press).

\bibitem{Lid97} Lidsey, J. E. et al., Abney, Rev. Mod. Phys., \textbf{69}, 373 (1997).

\bibitem{Lyt09} Lyth, D. H. \& Liddle, A. R.
2009, \emph{The Primordial Density Perturbation} (Cambridge
University Press).

\bibitem{Mal03} Maldacena, J. M., JHEP, \textbf{0305}, 013 (2003).

\bibitem{Bar04} Bartolo, N., Komatsu, E., Matarrese, S. \& Riotto, A., Phys. Rept., \textbf{402}, 103 (2004).

\bibitem{Che10} Chen, X., Adv. Astron., \textbf{2010}, 638979 (2010).

\bibitem{Fel11a} De Felice, A. \& Tsujikawa, S., Phys. Rev. D, \textbf{84}, 083504 (2011).

\bibitem{Fel11b} De Felice, A. \& Tsujikawa, S., JCAP, \textbf{1104}, 029 (2011).

\bibitem{Noz13a} Nozari, K. \& Rashidi, N., Phys. Rev. D, \textbf{88}, 023519 (2013).

\bibitem{Noz15} Nozari, K. \& Rashidi, N.,  Advances in High Energy Physics, https://doi.org/10.1155/2016/1252689 (2016).

\bibitem{Noz16a} Nozari, K. \& Rashidi, N., Physical Review D, \textbf{93},
124022 (2016).

\bibitem{Noz18a} Rashidi, N. \& Nozari, K., International Journal of Modern Physics D, \textbf{27}, 1850076 (2018).

\bibitem{Noz18b} Nozari, K. \& Rashidi, N., The Astrophysical Journal \textbf{863}, 133 (2018).

\bibitem{Noz19} Nozari, K. \& Rashidi, N., The Astrophysical Journal, \textbf{882}, 78 (2019).

\bibitem{Ras20} Rashidi, N. \& Nozari, K., The Astrophysical Journal, \textbf{890}, 58 (2020).

\bibitem{NoJ11} Nojiri, S. \& Odintsov, S. D, Phys. Rept. \textbf{505}, 59-144 (2011).

\bibitem{NoJ18} Nojiri, S., Odintsov, S.D. \$ Oikonomou, V.K., Phys.Rept., \textbf{692}, 1-104 (2017).

\bibitem{Sot10} Sotiriou, T. P., \& Faraoni, V., Rev. Mod. Phys \textbf{82}, 451 (2010).

\bibitem{Noj11a} Nojiri, S., \& Odintsov, S. D., Phys. Rept \textbf{505}, 59 (2011).

\bibitem{Sta07} Starobinsky, A. A., JETP Lett \textbf{86}, 157 (2007).

\bibitem{Noj16d} Nojiri, S., \& Odintsov, S. D., \& Oikonomou, V. K., Phys. Rev. D \textbf{93}, 084050 (2016).

\bibitem{Noj16e} Nojiri, S., \& Odintsov, S. D., \& Oikonomou, V. K., Mod. Phys. Lett. A \textbf{31}, 1650172 (2016).

\bibitem{Noj16f} Nojiri, S., \& Odintsov, S. D., \& Oikonomou, V. K., JCAP, \textbf{1605}, 046 (2016).

\bibitem{Fel10} Felice, A. De., \& Tsujikawa, S., Living Rev. Rel \textbf{13}, (2010).

\bibitem{Noj17b} Nojiri, S., \& Odintsov, S. D., \& Oikonomou, V. K., Phys. Rept \textbf{692}, 1 (2017).

\bibitem{Son07} Song, Y. S., \& Hu, W., \& Sawicki, I., Phys. Rev. D \textbf{75}, 044004 (2007).

\bibitem{Noj16c} Nojiri, S., \& Odintsov, S. D., \& Oikonomou, V. K., Class. Quantum. Grav \textbf{33}, 125017 (2016).

\bibitem{Noj19a} Nojiri, S., Odintsov, S.D., \& Oikonomou, V.K., DOI: 10.1016/j.nuclphysb.2019.02.008 (2019).

\bibitem{Odi19} Odintsov, S.D., \& Oikonomou, V.K., Phys. Rev. D \textbf{99}, 064049 (2019).

\bibitem{Bud17} Budhi, R. H. S., DOI: https://doi.org/10.1088/1742-6596/1127/1/012018 (2017).

\bibitem{Ber07} Bertolami, O., Boehmer, C. G., Harko, T., \& Lobo, F. S. N., Phys. Rev. D \textbf{75}, 104016 (2007).

\bibitem{Har08} Harko, T., Phys. Lett. B \textbf{669}, 376 (2008).

\bibitem{Tha11} Thakur, S., Sen, A. A., \& Seshadri, T. R., Phys. Lett. B \textbf{696}, 309 (2011).

\bibitem{Ber08a} Bertolamim O., \& Paramos, J., Class. Quantum Grav. \textbf{25}, 5017 (2008).

\bibitem{Boe08} Boehmer, C. G., Harko, T., \& Lobo, F. S. N., Astropart. Phys. \textbf{29}, 386 (2008).

\bibitem{Far09} Faraoni, V., Phys. Rev. D \textbf{80}, 124040 (2009).

\bibitem{Har10a} Harko,T., \& Lobo, F. S. N., The European Physical Journal C-Particles and Fields \textbf{70}, 373 (2010).

\bibitem{Noj04} Nojiri, S., \& Odintsov, S. D., Phys. Lett. B \textbf{599}, 137 (2004).

\bibitem{All05} Allemandi, G., Borowiec, A., Francaviglia, M., \& Odintsov, S. D., Phys. Rev. D \textbf{72}, 063505 (2005).

\bibitem{Ber08b} Bertolami, O., \& Paramos, J., Phys. Rev. D \textbf{77}, 084018 (2008).

\bibitem{Far07} Faraoni, V., Phys. Rev. D \textbf{76}, 127501 (2007).

\bibitem{Har10b} Harko, T., Phys. Rev. D \textbf{81}, 044021 (2010).

\bibitem{Raj17} Rajabi, F., \& K. Nozari, Phys. Rev. D \textbf{96}, 084061 (2017).

\bibitem{Har11} Harko, T., Lobo, F. S.N., Nojiri, S., \& Odintsov, S D., Phys. Rev. D \textbf{84}, 024020 (2011).

\bibitem{Sah17} Sahu, S. K., Tripathy, S. K., Sahoo, P. K., \& Nath, A., Chinese Journal of Physics, \textbf{55}, 862-869 (2017).

\bibitem{Wu18} Wu, J., Li, G., Harko, T., \& Liang, S.-D., The European Physical Journal C \textbf{78}, 430 (2018).

\bibitem{Paw19}Pawar, D .D.,  \& Shahare, S. P., https://doi.org/10.1016/j.newast.2019.101318 (2019).

\bibitem{Ros16} Roshan, M., \& Shojai, F., Phys. Rev. D, \textbf{94}, 044002 (2016).

\bibitem{Ak18a} Planck Collaboration: Aghanim, N., et. al. (2018) [arXiv:1807.06209[astro-ph.CO]].

\bibitem{Ak18b} Planck Collaboration: Akrami, Y., et. al. (2018) [arXiv:1807.06209[astro-ph.CO]].

\bibitem{Ak18c} Planck Collaboration: Akrami, Y., et. al. (2018) [arXiv:1807.06211[astro-ph.CO]].

\bibitem{Boa17}  Board, C. V. R. \& J. D. Barrow, Phys. Rev. D 96, no.12, 123517 (2017).

\bibitem{Nar18} Nari, N., \& Roshan, M., Phys. Rev. D \textbf{98}, 024031 (2018).

\bibitem{Sot02} Sotiriou, T. P., \& Faraoni, V., Class. Quant. Grav. \textbf{25}, 205002 (2008).

\bibitem{Ber08} Bertolami, O., \& Lobo, F. S. N., \& Paramos, J., Phys. Rev. D \textbf{78}, 064036 (2008).

\bibitem{Pol03} Polyanin, A. D. \& Zaitsev, V. F., 2003, \emph{Handbook of Exact Solutions for Ordinary Differential Equations},
(2nd Edition , Chapman \& Hall/CRC, Boca Raton).

\bibitem{Mar14} Martin, J., Ringeval, C., \& Vennin, V., Phys. Dark Univ. \textbf{5-6}, 75 (2014).

\bibitem{Lea02} Leach, S. M., Liddle, A. R., Martin, J., \&
Schwarz, D. J., Phys. Rev. D \textbf{66}, 023515 (2002).

\bibitem{Bam14} Bamba, K., Nojiri, S., Odintsov, S. D., \&
Saez-Gomez, D., Phys. Rev. D \textbf{90}, 124061 (2014).

\bibitem{Bha20} Bhattacharjee, S., Santos, J.R.L., Moraes,
P.H.R.S., \& Sahoo, P.K., Eur. Phys. J. Plus \textbf{135}, 576
(2020).




\end{thebibliography}
\end{document}